\documentclass{ws-ijmpa}
\usepackage{graphicx}
\usepackage{mathtools}
\usepackage{caption}
\usepackage[sorting=none, backend=biber]{biblatex}
\newcommand{\kaon}{\mathrm{K}^0}
\newcommand{\akaon}{\bar{\mathrm{K}}^0}

\newcommand{\Ks}{\mathrm{K_S}}
\newcommand{\Kl}{\mathrm{K_L}}

\newcommand{\Km}{\mathrm{K_{-}}}

\newcommand{\alktev}{A_L = (3.322\pm 0.058_{\textrm{stat}} \pm 0.047_{\textrm{syst}}) \times 10^{-3}}

\newcommand{\KsAs}{-4.9}
\newcommand{\KsAsErrStat}{5.7}
\newcommand{\KsAsErrSyst}{2.6}
\newcommand{\asresult}{A_S = (\KsAs \pm \KsAsErrStat_{\textrm{stat}} \pm \KsAsErrSyst_{\textrm{syst}}) \times 10^{-3}}
\newcommand{\asresultcombined}{A_S = (-3.8 \pm 5.0_{\textrm{stat}} \pm 2.6_{\textrm{syst}}) \times 10^{-3}}

\usepackage{mathtools}

\DeclarePairedDelimiter\ket{\lvert}{\rangle}
\DeclarePairedDelimiterX\braket[2]{\langle}{\rangle}{#1 \delimsize\vert #2}

\begin{document}
\markboth{Wojciech Krzemien, Elena Perez del Rio}{The KLOE-2 experiment: overview of the recent KLOE results}

%
\catchline{}{}{}{}{}
%

\title{The KLOE-2 experiment: overview of recent results}

\author{Wojciech Krzemien}
\address{High Energy Physics Division, National Centre for Nuclear Research, \\
 05-400 Otwock-\'Swierk, Poland\\
 wojciech.krzemien@ncbj.gov.pl}

\author{Elena P{\'e}rez del R{\'i}o}

\address{LNF-INFN, Laboratori Nazionali di Frascati,\\
Via E. Fermi 40, 00044, Frascati (RM), Italy\\
eperez@lnf.infn.it}
\author{on behalf of the KLOE-2 Collaboration}

\maketitle

\begin{history}
\received{Day Month Year}
\revised{Day Month Year}
\end{history}

\begin{abstract}
The KLOE detector at the DA$\Phi$NE $\phi$-factory has been operating in two periods from 2001 to 2006 and 
from 2014 to 2018 collecting a large sample of $\phi$-meson decays. This allowed to perform precision measurements and  studies of fundamental symmetries, and searches of New Physics phenomena.
In this overview, the results of KLOE and KLOE-2 Collaborations are presented. The most recent results from the KLOE experiment are discussed, covering: the measurement of the running fine-structure constant $\alpha_{em}$, the Dalitz plot measurement of $\eta \rightarrow \pi^+\pi^-\pi^0$, the search of a U boson, tests of discrete symmetries and quantum decoherence.
\end{abstract}


\ccode{PACS numbers:13.25.Es,13.40.Gp, 11.30.-j, 11.30.Cp }


\section{Introduction}

The KLOE (K LOng Experiment) experiment and its continuation KLOE-2, at the DA$\Phi$NE $\phi$-factory at the Laboratori Nazionali di Frascati in Italy, is a low-energy and high-precision experiment able to measure mesons in the range of energy from 5 MeV to 1 GeV.
Its program addresses various topics of modern physics~\cite{physics_kloe2} including
the study of the Standard Model (SM) dynamics, dark matter mediator searches, 
tests of quantum mechanics (QM) and discrete symmetries in neutral kaon systems.

The DA$\Phi$NE facility is a $e^{+} e^{-}$ collider that operates at a center of mass energy adjusted to the $\phi$ meson mass ($m_{\phi} \approx 1.019\, \text{GeV}$),
which decays predominantly to charged (49\%) and neutral kaons (34\%) , and to the  $\rho\pi$, $\pi^+\pi^-\pi^0$ (15\%) and $\eta\gamma$ (1.3\%) final states.
Being a $e^{+}e^{-}$ collider, DA$\Phi$NE provides a clean experimental environment, avoiding lots of strong-interaction backgrounds present in the hadronic accelerators. The initial-state energy is well known, allowing a precise calculation of the kinematics of the events, making the final states easy to reconstruct by looking at the event topologies. 
Also, DA$\Phi$NE, working at a fixed center-of-mass energy, allows to scan a broad energy range for measurements of the hadronic cross section or, e.g., dark mediator searches, exploiting events where one of the electrons (positrons) in the initial state radiated a photon, called initial-state radiation (ISR) events. For the ISR events, the invariant mass $\rm{M_{hadr}}$ of the hadronic system is reduced, thus, giving access to an energy range $\rm{M_{hadr}} < \sqrt{s}$.

For kaon physics, the main experimental advantage is the possibility to select a pure, kinematically well-defined $K_{S}$ beam. The $\phi$ particle is a vector meson with well-defined quantum numbers $J^{PC} = 1^{--}$, decaying via the strong, parity-conserving process $\phi \rightarrow \Kl \Ks$, and at DA$\Phi$NE those neutral kaons are produced as almost collinear, monochromatic pairs with longitudinal momenta of  100 MeV/c and transverse momenta of 15 MeV/c. The reconstruction of the $\Kl$ particles, with larger decay times, interacting in the calorimeter allows to efficiently tag the $K_{S}$ beam. Finally, at DA$\Phi$NE it is possible to study the entangled quantum system formed by the correlated neutral kaons with high statistics and in a low-background environment. The entangled pairs can be used to study a variety of interferometric patterns analogical to the one observed for entangled photons. Those phenomena, being a manifestation of quantum interference, provide access to the studies of the QM basics, as well as to precise CP, T and CPT symmetry tests.

Following the construction of the collider in 1989, the first KLOE experiment~\cite{DAFNE-KLOE} run from 2001 to 2006, acquiring a total integrated luminosity of $2.5\,\mathrm{fb^{-1}}$ at the $\phi$ peak, $\sqrt{s} = 1.019 \, \text{GeV}$, and about $250\,\mathrm{pb^{-1}}$ off-peak.
In 2008, the DA$\Phi$NE facility underwent an upgrade which included a new interaction scheme~\cite{DAFNE-2}. At the same time, new detectors were added to the KLOE experiment to face a new physics program ~\cite{DAFNE-KLOE2,KLOE2_proposal}.
Starting in 2014 and until March 2018, the new KLOE-2 experiment has been collecting data. In this period, a sample of $5.5\,\mathrm{fb^{-1}}$ has been acquired. Together with the KLOE data sample, the total collected data sets correspond to an integrated luminosity of $8\,\mathrm{fb^{-1}}$, which with about $2.4\times10^{10}$ $\phi$-mesons recorded, represents the largest sample acquired at a $\phi$-factory at the moment.

In this contribution, we present the recent results from the KLOE-2 Collaboration.
The article is structured in the following way. In sections~\ref{sec:daphne}  and~\ref{sec:detectors} the DA$\Phi$NE collider and the KLOE-2 experiment are described, respectively. Section~\ref{sec:hadron_phys} is dedicated to the results of the KLOE-2 Collaboration on hadron physics. 
Section~\ref{sec:dark_matter} covers the dark mediator searches performed in the last years and section~\ref{sec:discrete_symmetries} introduces results on kaon physics for discrete symmetries and quantum decoherence tests. Finally, in section~\ref{sec:outlook} an outlook and prospects are presented.

\section{The DA$\Phi$NE $\phi$-factory}
\label{sec:daphne}
\begin{figure}[htb!]
  \centerline{\includegraphics[width=10cm]{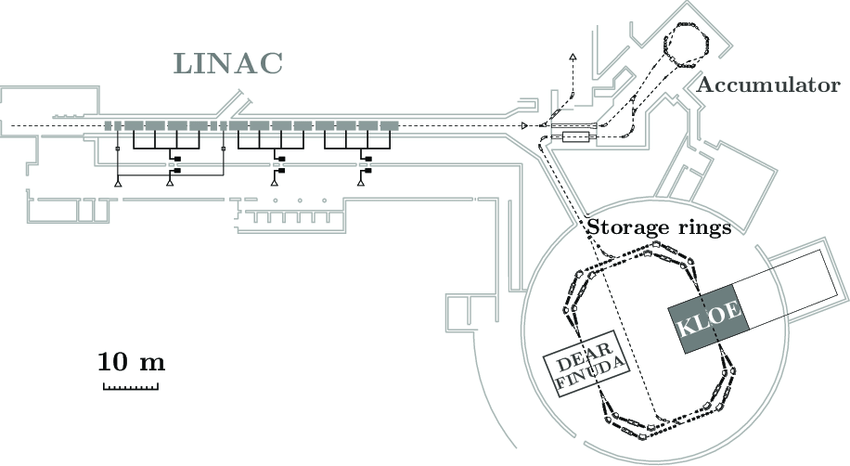}}
  \caption{DA$\Phi$NE complex schematic view. In the picture,the LInear ACcelerator (LINAC) section, followed by the accumulator ring, and the storage ring, with two separated IP regions where the KLOE and FINUDA experiments are located. Figure adapted from~\cite{nuovocimento}.}
  \label{fig:dafne_scheme}
\end{figure}

The DA$\Phi$NE complex is a $e^+e^-$ collider composed of two rings, one for each leptonic charge, providing collisions at the mass of the $\phi$-resonance (1.019 GeV), and an injection system. Each ring is $\sim 97 \,\text{m}$ long and they meet in a common interaction point (IP), where the experimental setup is placed. The injection system, which comprises an S-band (2856 MHz / 2998 MHz) linear accelerator (LINAC), 180 m long transfer lines and an accumulator/damping ring (see Fig.~\ref{fig:dafne_scheme}), is capable of providing fast and high-efficiency electron and positron injections.

In order to increase the instantaneous luminosity in collision mode, a new approach to the beam-beam interaction, a {\it Crab-Waist}~\cite{CW1, CW2,Milardi:eeFACT2018-MOYAA02} collision scheme, was designed, installed and commissioned with the SIDDHARTA~\cite{siddharta, DAFNE-2} setup. 
After a successful campaign with the KLOE-2 detector, which lasted from 2014 to 2018, the complex is foreseen to start operation with the upgraded SIDDHARTA-2 experiment~\cite{siddharta2}.

\section{KLOE-2 detectors}
\label{sec:detectors}
\begin{figure}[htb]
  \centerline{\includegraphics[width=6cm]{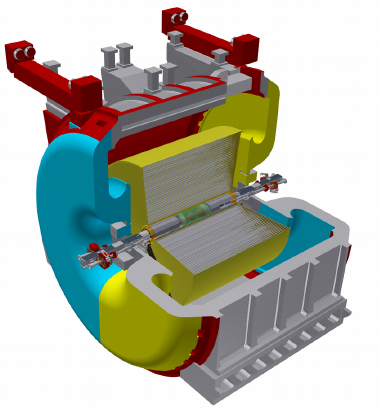}}
  \caption{KLOE-2 detector schematic view. From inner most the figure depicts the IP sphere surrounded by the Inner Tracker (translucent green), the QCALT calorimeters on the sides of the Inner Tracker, the DC (enclosed by yellow) and the closing end-caps (solid blue and yellow). The ECAL surrounding the DC is not shown.}
  \label{fig:kloe_layout}
\end{figure}

The KLOE-2 setup is a 4$\pi$ detector, which consists of two distinct assembles, the detector arrangement of the original KLOE experiment and additional tracker, calorimeters and taggers for $\gamma\gamma$-physics, which run in the 2014-2018 campaign. The schematic view of the KLOE-2 detector is presented in Fig.~\ref{fig:kloe_layout}.

The main parts of the KLOE-2 detector are a large drift chamber (DC)~\cite{driftchamber}, which allows to track the charge particles along the volume of the detector, and a lead scintillator electromagnetic calorimeter (ECAL)~\cite{calorimeter}, which provides energy and time measurements of particles interacting electromagnetically within the detector, all embedded in a super-conductive solenoid providing a $0.52 \,\text{T}$ magnetic field. The DC, largest of its type in the world, has a 4 m diameter and is 3.3 m long. It is built out of carbon fiber composite and consists of 12582 drift cells made of tungsten sense wires. A gas mixture of helium (90\%) and isobutane (10\%) fills the chamber volume. The position resolutions are $\sigma_{xy} \approx 150\,\mu \text{m}$ and $\sigma_{z} \approx 2\,\text{mm}$, and the momentum resolution, $\frac{\sigma_{p\perp}}{p\perp}$, is better than 0.4\% for large-angle tracks.
The ECAL covers $98\%$ of the solid angle and it is divided in a barrel and two side detectors made of a total of 88 modules. Each module is built out of 1 mm diameter scintillating fibers grouped in cells of $4.4 \times 4.4 \,\text{cm}^{2}$ and embedded in 0.5 mm lead foils; the readout is made of photo-multipliers from both sides. ECAL provides good energy resolution and excellent timing performance, with $\frac{\sigma(E)}{E} = \frac{5.7\%}{\sqrt{E\,(\text{GeV})}} $ and $\sigma(t) =\frac{ 57\text{ps}}{\sqrt{E\,(\text{GeV})}} \oplus 140 \,\text{ps}$ respectively.

The KLOE-2 upgrade completed this setup by adding new detectors to the original layout. Around the IP, a tracker device, the Inner Tracker (IT) was added. The IT consists of four concentric cylindrical gas electron multiplier detector (CGEM) layers~\cite{cgem-it} between the IP and the DC to improve the resolution on decay vertices close to the IP. Each layer consists of a triple-CGEM detector with an X-V strip readout. The X strips are placed longitudinally along the Z beam axis while the V strips have on average a 23-degree angle with respect to the beam axis. Both, X and V strips, have a $650\,\mu \text{m}$ pitch. The CGEMs are filled with a $Ar:{\it i}C_{4}H_{10}$ gas mixture. A technology with low material budget (below $2\%$ of radiation length $X_{0}$, altogether) was chosen, to minimize multiple scattering of low-momentum tracks, photon conversions and kaon regeneration.

For the $\gamma\gamma$-physics program of KLOE-2, the detector was equipped with two tagger systems, the Low-Energy Taggers (LETs) and the High-Energy Taggers (HETs), in order to detect electrons and positrons scattered in $e^{+} e^{-} \rightarrow e^{+}e^{-}\gamma^{*}\gamma^{*} \rightarrow e^{+}e^{-}X$ reactions. The HET~\cite{HET-LET} consists of two identical stations positioned at 11 m from the IP, to tag the $e^{\pm}$ with energies greater than 420 MeV, and escaping the beam orbit. Each HET detector consists of a sensitive area made up of 28 plastic scintillators with dimensions $3 \times 5 \times 6 \,\text{mm}^{3}$, read out by means of photo-multipliers. On the other hand, the LET~\cite{HET-LET} consists also of two identical stations, this time located  $1\,\text{m}$ right after the IP, in order to tag electrons and positrons with energy $160 < E < 400 \,\text{MeV}$. The two stations are built of an array of 54 LYSO crystals read out by a Silicon Photo-multipliers (SiPM).

Finally, two new calorimeters were added to the setup: The Quadrupole CALorimeters with Tiles (QCALT)~\cite{QCALT}, with two identical $1\,\text{m}$ long sample calorimeter, which are positioned at both sides of the IP around the DA$\Phi$NE low-$\beta$ quadrupoles. Each one is made of 5 layers of $5\,\text{mm}$ thick scintillator plates, alternated with $3.5\,\text{mm}$ thick tungsten plates, for a total of $\sim 5 X_{0}$, optically connected to the SiPM. The QCALT detects photons emitted at low angle, close to the interaction point (down to $10^\circ$). And the Crystal CALorimeters with Timing (CCALT)~\cite{CCALT}, which is built of LYSO crystals, and the light produced by the particles is collected by SiPMs. The QCALT instruments the quadrupoles at the beam lines of DA$\Phi$NE and it provides coverage for $K_{L}$ decays in the quadrupole region.

\section{Recent results on hadron physics}
\label{sec:hadron_phys}
\subsection{Running of the fine-structure constant $\alpha_{em}$ below 1 GeV}
The effective QED constant, $\alpha_{em}$, is known~\cite{arbuzov, abbiendi, acciarri, odaka, levine} to increase with rising momentum transfer due to vacuum polarization (VP) effects. The VP effects can be implemented by making the fine-structure constant explicitly $s$-dependent ($s=q^{2}$): 
\begin{equation}
\alpha_{em}(q^2) = \frac {\alpha_{em}(0)}{1 - \Delta\alpha(q^2)}.
\end{equation}
The correction $\Delta\alpha$ represents the sum of the lepton, the five lightest quarks, and the top quark contributions:
\begin{equation}
\Delta\alpha(q^2) = \Delta\alpha_{lep}(q^2) + \Delta\alpha^{(5)}_{had}(q^2) +\Delta\alpha_{top}(q^2). 
\end{equation}
At low energies the top quark contribution is negligible, since $\Delta\alpha_{top} \propto \frac{s}{m^{2}_{top}}$. The leptonic part is calculated in perturbation theory up to the third order, at which it takes the value $\Delta\alpha_{lep}(q^2 = M^{2}_{Z}) = 314.98 \times 10^{-4}$, with $M_{Z} \approx 91.19 \,\mathrm{GeV}$~\cite{kallen, steinhausen}. The hadronic part cannot be calculated with perturbative methods. Instead, it can be evaluated from experimental data by means of the dispersion relation :
\begin{equation}
\Delta\alpha^{(5)}_{had}(s) = - \frac{\alpha(0) s}{3\pi} \int_{s0}^{\infty} \frac{R_{had}(s')}{s'(s'-s-i\varepsilon)}ds' ,
\end{equation}
where  $R_{had}(s) = \frac{(e^+e^- \rightarrow \mathrm{hadrons})}{(e^+e^- \rightarrow \mu^+\mu^
-)}$.

The value of $\alpha_{em}(s)$ in the time-like region can be extracted from the ratio of the differential cross-section of $e^+e^- \rightarrow \mu^+\mu^-\gamma$, with a photon from the Initial State Radiation 
(ISR), and the corresponding cross-section obtained from the Monte Carlo (MC) simulation with $\alpha_{em}(s) = \alpha_{em}(0)$~\cite{aem-1}.
\begin{equation}
  \centering
  \left|\frac{\alpha_{em}(s)}{\alpha_{em}(0)}\right|^2 = \frac{d\sigma_{data}(e^+e^- \rightarrow \mu^+ \mu^- \gamma (\gamma))|_{ISR}/d\sqrt{s}}{d\sigma^0_{MC}(e^+e^- \rightarrow \mu^+ \mu^- \gamma (\gamma))|_{ISR}/d\sqrt{s}}  
  \label{eq:1}
\end{equation}

\begin{figure}[htb!]
  \centerline{\includegraphics[width=10cm]{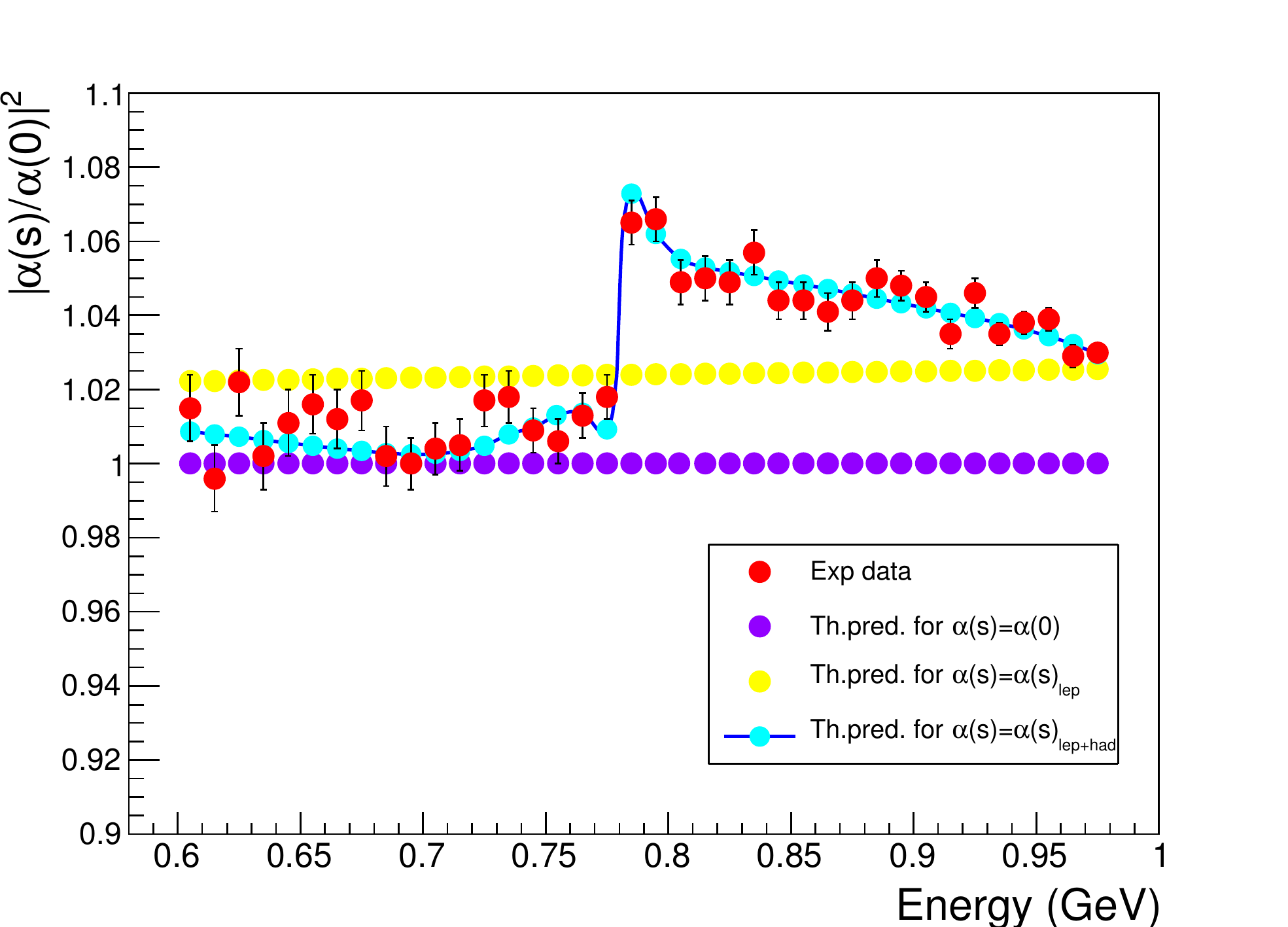}}
  \caption{ $\left|\frac{\alpha_{em}(s)}{\alpha_{em}(0)}\right|^2$ as a function of the di-muon invariant mass compared with the prediction. The red points are the KLOE data with statistical errors; the violet points correspond to the theoretical prediction for a fixed coupling $(\alpha(s) =\alpha(0))$; yellow points are the prediction with only virtual lepton pairs contributing to the shift $\Delta \alpha(s) = \Delta \alpha (s)_{lep}$, and the cyan points with the solid line are the full QED prediction with both lepton and quark pairs contributing to the shift $\Delta \alpha (s)= \Delta \alpha (s)_{lep+had}$.}
  \label{fig:aem1}
\end{figure}

In Fig.~\ref{fig:aem1} the ratio (\ref{eq:1}) is compared to the theoretical predictions~\cite{aem_theory}.
In the time-like region of $q^2$, $\Delta\alpha$ is a complex quantity and the real part of it expressed as:

\begin{equation}
  \centering
  \Re\Delta\alpha= 1 - \sqrt{|\alpha_{em}(0)/\alpha_{em}(s)|^2 - (\Im\Delta\alpha)^2}
  \label{eq:aem2}
\end{equation}

The imaginary part contribution is usually neglected when calculating the running of $\alpha$. In general, this is a good approximation since this contribution enters at order $\mathcal{O}(\alpha^{2})$ while the real part enters the calculations at order $\mathcal{O}(\alpha)$. Nevertheless, in the region of the $\rho$ meson resonance, which cross section can be measured with an accuracy better than 1\%, the $\Im\Delta\alpha$, should be taken into account.

\begin{figure}[htb!]
  \centering
  \includegraphics[width=.8\linewidth]{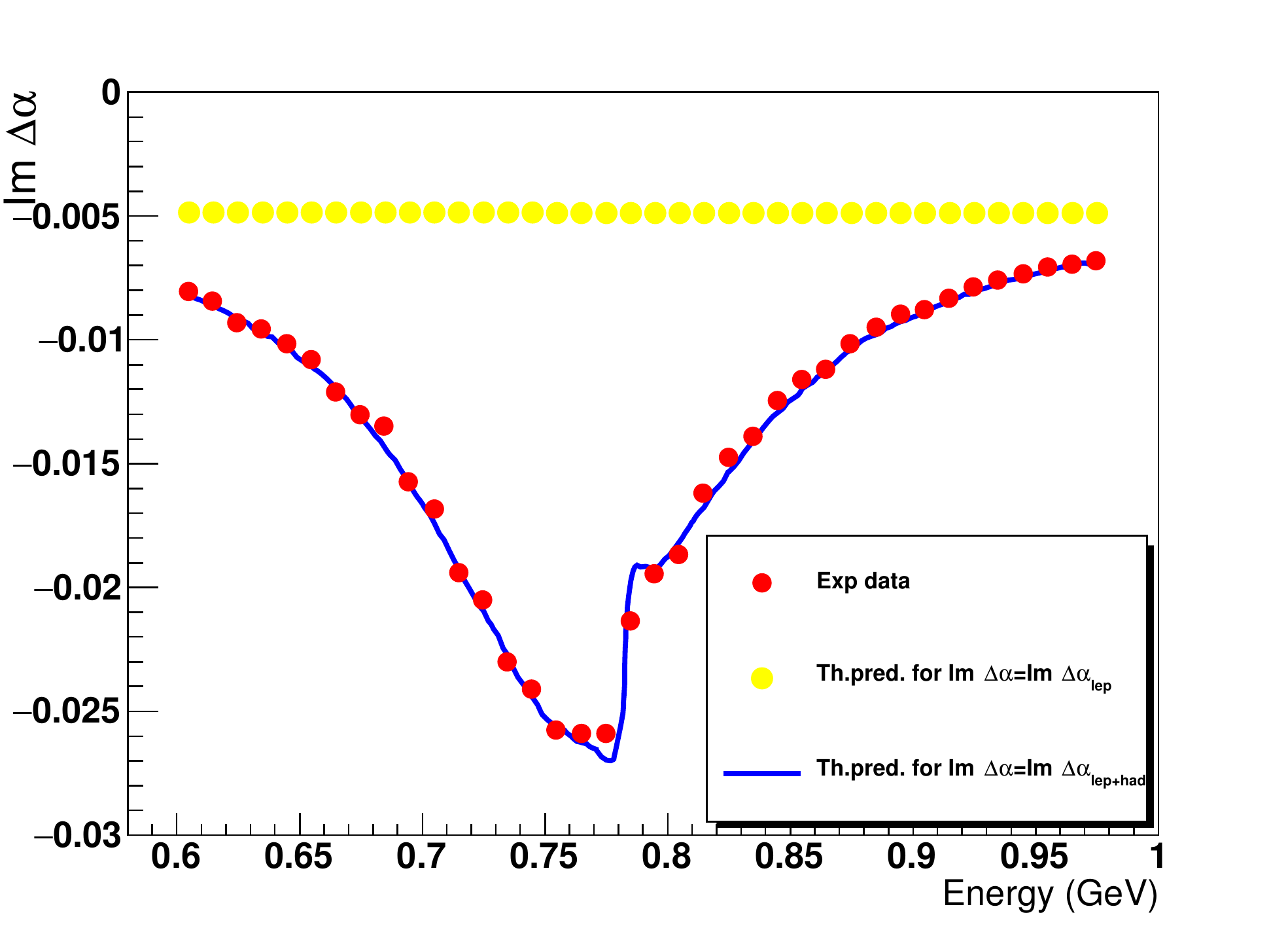}
  \includegraphics[width=.85\linewidth]{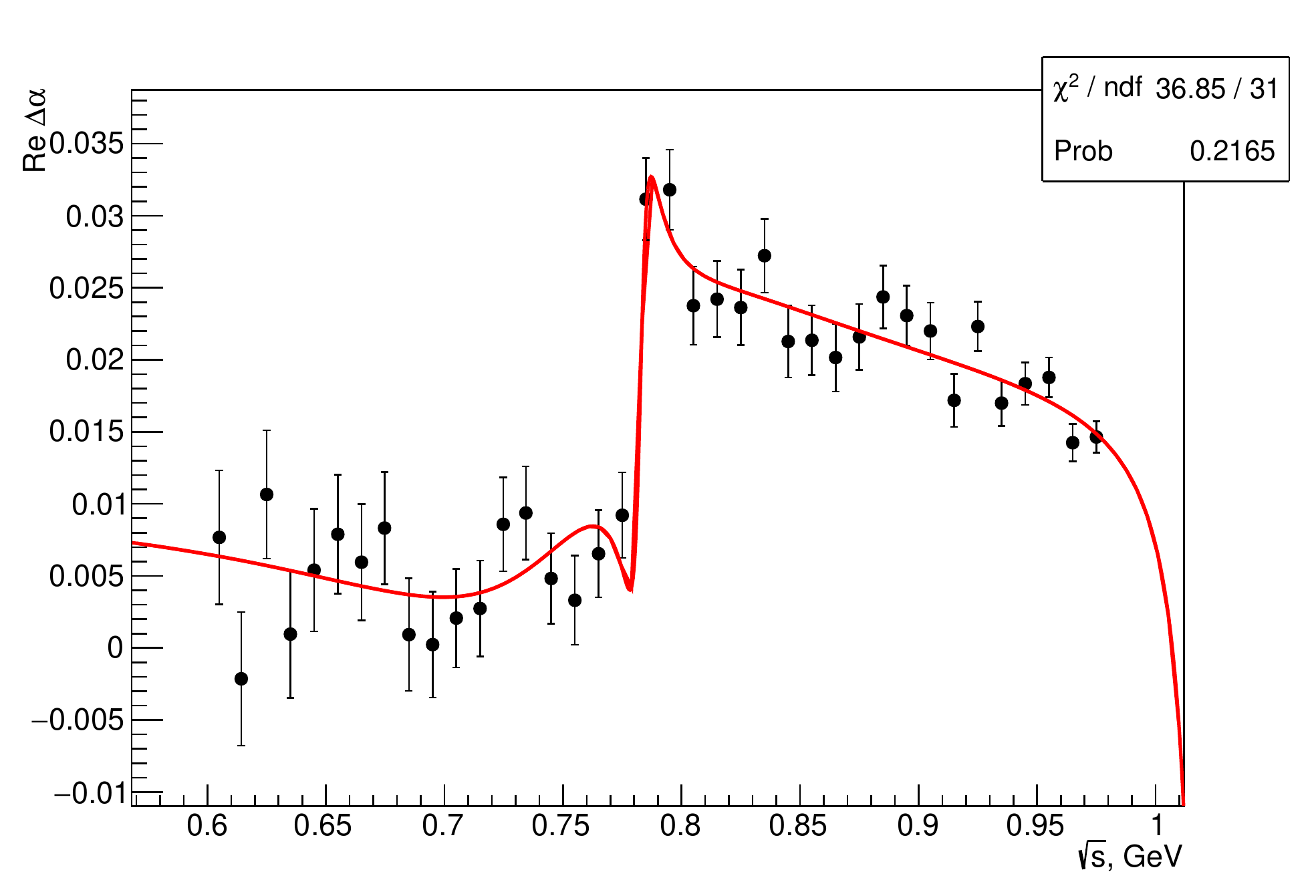}
  \caption{Up: $\Im \Delta\alpha$ from KLOE $\pi^+\pi^-\gamma$ cross section (red points) compared to the predictions from a compilation of other measurements for $\Im \Delta \alpha = \Im \Delta \alpha_{lep}$ (yellow points) and $\Im \Delta \alpha= \Im \Delta \alpha_{lep+had}$ only for $\pi\pi$ channels (blue solid line); Down: $\Re\Delta\alpha$ from Eq.~\ref{eq:aem2} (black points), the red line is the fit described in the text.}
  \label{fig:aem2}
\end{figure}

From the optical theorem follows that $\Im\Delta\alpha = -\frac{\alpha_{em}}{3}R_{had}(s)$, and the upper panel in Fig.~\ref{fig:aem2} presents the behavior of $\Im\Delta\alpha$ obtained from the KLOE data on the $\pi^+\pi^-$ cross-section
~\cite{2picross}. The real part from Eq.~\ref{eq:aem2} is shown in the lower panel of Fig~\ref{fig:aem2}. The superimposed curve is a fit performed by parameterizing the $\omega$(782) and $\Phi$(1020) resonances using Breit-Wigner distributions and describing the $\rho$(770) form-factor with the Gounaris-Sakurai parameterization~\cite{G-S}, and adding a non-resonant term. The mass
 and the width of the $\Phi$, the $\omega$ width as well as the product Br($\Phi \rightarrow e^+e^-$)Br($ \Phi \rightarrow \mu^+\mu^-$) have been fixed to the PDG values~\cite{PDG}. Assuming lepton universality and correcting for phase space, the value Br($\omega \rightarrow \mu^+\mu^-$)=($6.6 \pm 1.4 \pm 1.7)\times10^{-5}$ is obtained \mbox{(PDG: Br($\omega \rightarrow \mu^+\mu^-$) = $(9.0 \pm 3.1)\times10^{-5})$}.
The results show more than 5$\sigma$ significance of the hadronic contribution.

\subsection{Precision measurement of the $\eta \rightarrow 3\pi$ Dalitz plot distribution}

The isospin-violating decay $\eta \rightarrow 3\pi$ is induced dominantly by the strong interaction via a term proportional to the differences of the masses of the  $u$ and $d$ quarks. Therefore, the decay is a perfect laboratory for testing chiral perturbation theory, ChPT~\cite
{ChPT}. The decay amplitude is proportional to $Q^{2}$, the quadratic quark mass ratio defined as $Q^2 = \frac{m_s^2 - \hat{m}^2}{m_d^2 - \hat{m}^2}$ where $\hat{m} = \frac{1}{2}\left(m_d+m_u \right)$. Thus, the precise experimental distributions could be used directly for the dispersive analysis to determine the $Q$ ratio.

\begin{figure}[htb!]
  \centering
  \includegraphics[width=.8\linewidth]{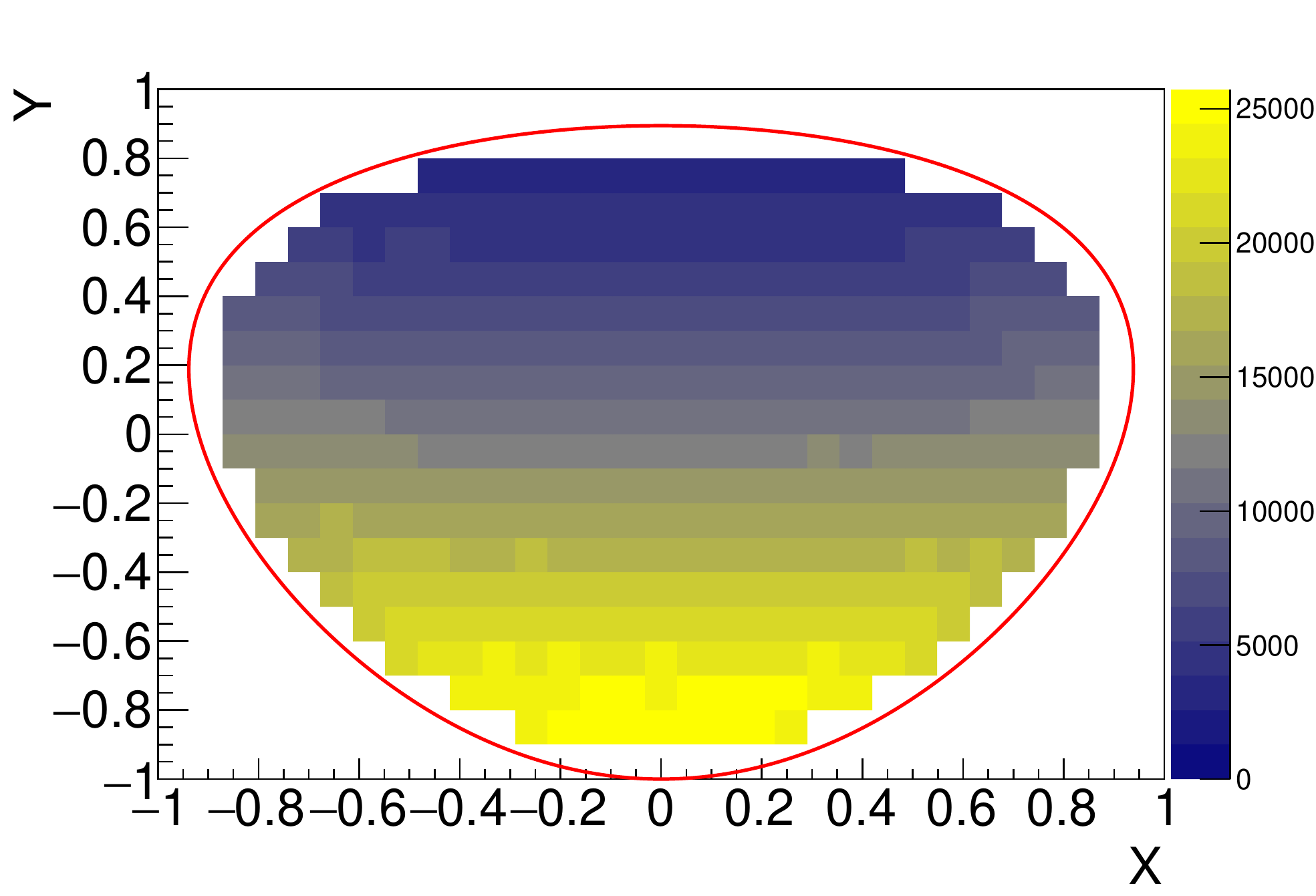}
  \caption{(Color online) The experimental background-subtracted Dalitz plot distribution represented by the two-dimensional histogram with 371 bins. The red line indicates the physical border while only bins used for the Dalitz parameter
fits are shown. Figure adapted from Ref.~\cite{KLOEdalitz_16}.}
  \label{fig:dalitz}
\end{figure} 

\begin{table}[htbp]
  \centering
  \resizebox{\textwidth}{!}{\begin{tabular}{|c|c|c|c|c|c|}
  \hline
   & a & b & d & f & g \\
  \hline
  KLOE('16)~\cite{KLOEdalitz_16} & $-1.095 \pm 0.004$ & $0.145 \pm 0.006$ & $0.081 \pm 0.007$ & $0.141 \pm 0.011$ & $-0.044 \pm 0.016$ \\
  KLOE('16)~\cite{KLOEdalitz_16} & $-1.104 \pm 0.004$ & $0.142 \pm 0.006$ & $0.073 \pm 0.005$ & $0.154 \pm 0.008$ &\\
  KLOE('08)~\cite{KLOEdalitz_08} & $-1.090 \pm 0.020$ & $0.124 \pm 0.012$ &  $0.057 \pm 0.017$ &  $0.14 \pm 0.02$ & \\
  WASA~\cite{wasa_dalitz} & $-1.144 \pm 0.018$ & $0.219 \pm 0.051$  & $0.086 \pm 0.023$ &  $0.115 \pm 0.037$ & \\
  BESIII~\cite{bes_dalitz} & $-0.128 \pm 0.017$ & $0.153 \pm 0.017$ & $0.085 \pm 0.018$ & $0.173 \pm 0.035$ &\\
  \hline
  \end{tabular}}
  \caption{The table presents the Taylor expansion parameters of the squared amplitude of the decay from the  KLOE('16) analysis compared to previous measurements.The $c$ and $e$ parameters must be zero assuming charge conjugation symmetry. The $g$ parameter has been extracted for the first time.}
  \label{table:dalitz}
\end{table}

The Dalitz plot distribution is expressed in terms of the normalized variables: $X = \sqrt{3} \frac{T_+T_-} {Q_{\eta}} $ and $Y = 3 \frac{T_0}{Q_{\eta}} - 1$ where the $T_i$ are the kinetic energies of the pions in the $\eta$ rest frame and $Q_{\eta} = m_{\eta}-2m_{\pi^{\pm}}-m_{\pi^0}$. The squared amplitude is usually parametrized as a Taylor expansion around the center, $|A(X;Y)|^2 \simeq N(1+ aY + bY^2 + cX + dX^2 + eXY + fY^3 + gX^2Y + ...)$.

The KLOE-2 Collaboration performed a measurement of the Dalitz plot by using the decay $e^+e^- \rightarrow \Phi \rightarrow \eta \gamma$, with $\eta \rightarrow \pi^+\pi^-\pi^0$, on a sample of 1.6 fb$^{-1}$ of data, corresponding to $4.7\times10^6$ events~\cite
{KLOEdalitz_16}, improving the previous measurement~\cite{KLOEdalitz_08} obtained by the KLOE Collaboration, and extracting the $g$ parameter of the Taylor expansion for the first time. The Dalitz plot is shown in Fig.~\ref{fig:dalitz} and a list of the extracted parameters, compared to other measurements, is presented in Table~\ref{table:dalitz}. The c and e parameters are C-violating and are compatible with zero.

\section{Dark matter searches}
\label{sec:dark_matter}
The SM, although being the most complete theoretical framework at present, does not provide a definitive model of all elementary particles. Some unexplained observations as the $511$ keV gamma-ray signal from the galactic center~\cite{integral}, the CoGeNT results~\cite{cogent}, the DAMA/LIBRA annual modulation~\cite{dama,libra}, the total $e^+e^-$ flux~\cite{atic, hess1, hess2, fermi} and the discrepancy between theory and experiment in the magnetic moment of the muon, $a_{\mu}$ have motivated possible explanations based on Physics Beyond the Standard Model (BSM).  Among them, some extensions of the SM~\cite{ext1, ext2, ext3, ext4,ext5} claim to explain the aforementioned anomalies by means of dark matter models, with a Weakly Interacting Massive Particle (WIMP) belonging to a secluded gauge sector or hidden symmetry, $U(1)_{DM}$. The new gauge interaction would be mediated by a vector gauge boson, called the U boson or dark photon, which could interact with the photon via a kinetic mixing term in the Lagrangian, which is a coupling between the $U(1)$ of the SM hypercharge and the $U(1)_{DM}$:
\begin{equation}                                  
\mathcal{L}_{mix} = - {\frac{\varepsilon}{2}} F^{EM}_{\mu\nu}{F^{DM}}^{\mu\nu}
\end{equation}
The $\varepsilon$ parameter represents the mixing strength and it is defined as the ratio of the dark to the SM electroweak coupling, $\alpha_{DM}/\alpha_{EM}$. The $F^{EM}_{\mu\nu}$ and $F^{DM}_{\mu\nu}$ are the field strengths of the $U(1)$ and $U(1)_{DM}$ dark photon field, respectively~\cite{ext4,ext1}. Via this coupling, the U boson would be allowed to decay not only to dark particles (invisible decays) but also to SM particles (visible decays).
A U boson, with mass of $\mathcal{O}(1 \,\text{GeV})$ and $\varepsilon$ in the range of $10^{-2} - 10^{-7}$, could be observed in $e^+e^-$ colliders via different processes: $e^+e^- \rightarrow U \gamma$, $V \rightarrow P\gamma$ decays, where $V$ and $P$ are vector and pseudoscalar mesons, and $e^+e^- \rightarrow h^{\prime}U$, where $h^{\prime}$ is a Higgs-like particle responsible for the breaking of the hidden symmetry. 

\begin{figure}[htb!]
\centering
\includegraphics[width=10cm]{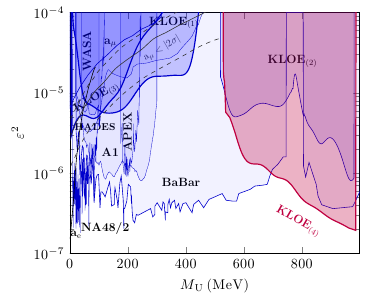}
\caption{90\% CL exclusion plot  for $\varepsilon^2$ as a function of the $\mathrm{U}$-boson mass. 
The solid lines are the limits from the muon and electron anomaly. The gray line shows the U boson parameters that could explain the discrepancy between SM prediction and the experimental value of muon anomalous magnetic momentum, $a_{\mu}$, with a $2
\, \sigma$ error band (gray dashed lines). The presented upper limits correspond to the $U \rightarrow e^{+} e^{-}$ limit from the $\phi \rightarrow \eta U$ dalitz decay (KLOE(1)), the $U \rightarrow e^{+} e^{-}$ limit (KLOE(3)) from radiative production, the $U \rightarrow \mu^{+} \mu^{-}$ constraint (KLOE(2)) and the combined limit from $U \rightarrow \mu^{+} \mu^{-}$,$\pi^{+} \pi^{-}$ at full KLOE statistics (KLOE(4)) in comparison with the competitive limits by
BaBar~\cite{babar}, NA48/2~\cite{na48} and LHCb experiments~\cite{lhcb}. 
}
\label{fig:upper_limit}       
\end{figure}

The KLOE-2 Collaboration searched for the signature of the dark photon by investigating the following three processes:\\
\textit{\bf $\phi$-Dalitz Decay:}\\
As explained previously, the dark photon is expected to be produced in vector to pseudoscalar meson decays, thus, producing a signature peak in the invariant mass distribution of the electron-positron pair over the continuum Dalitz background. 
The KLOE-2 Collaboration searched for the U boson by analyzing the $\phi \rightarrow \eta \mathrm{e}^+ \mathrm{e}^-$ decay, where the $\eta$ meson is tagged by $\pi^+ \pi^- \pi^0$~\cite{KLOE_UL1}, using an integrated luminosity of 1.5~fb$^{-1}$, and also in the $3 \pi^0$ decays~\cite{KLOE_UL2}, for which a total integrated luminosity of 1.7~fb$^{-1}$ was employed. 
The final combined limit is shown in Fig.~\ref{fig:upper_limit} and dubbed as KLOE$_{(1)}$.
This limit~\cite{KLOE_UL2} rules out a wide range of U boson parameters that could explain the $a_{\mu}$ discrepancy in the hypothesis of a visibly-decaying dark photon.\\ 
\textit{\bf Higgsstrahlung process:}\\
 In analogy to the SM case, since the U boson needs to be massive, one can implement a spontaneous symmetry breaking mechanism of the $U(1)_{DM}$ symmetry, introducing a Higgs-like, $h^{\prime}$ particle, or {\it dark Higgs}. The mass hierarchy between the $h^{\prime}$ and the U boson is not constrained by the theory~\cite{Batell}. The KLOE-2 Collaboration investigated also the Higgsstrahlung process, sensitive to the dark coupling constant $\alpha_{\rm DM}$. The invisible scenario, where the dark Higgs is lighter than the U boson and escapes detection was considered~\cite{enrico}. In this case, the expected signal is a muon pair from the U boson decay plus missing energy.
The analysis was performed using two data samples: one collected on the $\phi$ peak ($\sqrt(s) = 1019 \,\rm MeV$) and a second one at $\sqrt{s}=1000$~MeV (off-peak sample), where all the backgrounds from the $\Phi$ decays are strongly suppressed.
 No signal signature was observed and a Bayesian limit on the number of signal events at 90\% CL was evaluated, bin-by-bin, for the on-peak and off-peak sample separately and the results were presented in terms of  $\alpha_\mathrm{DM} \times \varepsilon^2$. 
Assuming the hypothesis of $\alpha_\mathrm{DM}=\alpha$, the upper  limits on the kinetic mixing parameter $\varepsilon$ ranges between $10^{-4}$ and $10^{-3}$.\\
\textit{\bf U boson radiative production:}\\
In KLOE-2 also both leptonic and hadronic decays of the U boson produced in $e^+e^-$ annihilation were investigated by studying the reactions: $ \rm U \gamma_{ISR}\to \mu^+ \mu^-,\, \pi^+\pi^-$ and $\rm U \gamma_{ISR} \to \mathrm{e^+ e^-}$, all of them by means of the Initial State Radiation Method.
The analysis of the  $\rm U \gamma_{ISR} \to \mathrm{e^+ e^-}$ reaction, on a sample of 1.5 fb$^{-1}$, profited of the high rate of radiative Bhabha events produced in KLOE and allowed to search in the lower mass region, setting limits down to $5 \,\rm MeV$ in the di-electron invariant mass, or U boson mass. In contrast, a search using the decay in muon-pairs was performed to explore the higher mass region.
 However, the result was updated by studying the decay of the U boson into pion-pair, which is of special interest in order to recover sensitivity in the mass range of the $\rho - \omega$ resonance region. About this mass, the mixing of the $\gamma$ to the $\rho$ meson makes the decay into hadrons more probable. For this reason, the searches for the U boson decaying into $\mu$-pairs have no sensitivity in this mass range. To achieve the desired reach, KLOE-2 re-analyzed the $\rm U\to \mu^+ \mu^-$ decay at full statistics together with the decay into $\pi^+\pi^-$. The combined upper limit was extracted for the mass region between 650 MeV and 987 MeV. 

For the combined search in $U\to \pi^+\pi^-, \mu^+\mu^-$ a total of $1.9 \,\rm fb^{-1}$ integrated luminosity were used in both channels. Since no dark photon signature was observed, limits at 90\% CL were extracted for both processes on the number of excluded U candidate events and combined in a common limit. The limits on the mass of U were converted into limits on the kinetic mixing parameter $\varepsilon^2$ by using the formula reported in Refs.~\cite{eeg,mmg,ppg}. These limits are shown in Fig.~\ref{fig:upper_limit} (KLOE$_{(2,3,4)}$) together with the results from other experiments at the time of the publication.

\section{Discrete symmetries and quantum decoherence tests}
\label{sec:discrete_symmetries}
Since weak interactions change the flavor, the full Hamiltonian describing the neutral meson states cannot commute with the flavor operator. Consequently, the Hamiltonian eigenstates, corresponding to physical states $K_{L}$ and $K_{S}$, with defined lifetimes, cannot be identified as the flavor eigenstates $\ket{K_{0}}$ and $\ket{\bar{K}_{0}}$. 
However, it is known that in kaons system the CP symmetry violation effect is small~\cite{PDG_CP}, therefore physical kaon states $K_{S}$ and $K_{L}$ are almost CP eigenstates $\ket{K_1}$ and $\ket{K_2}$, respectively. The admixture of the second CP state is quantified by introducing (small) complex parameters $\epsilon_S$ and $\epsilon_L$~\cite{cp_sozzi}: 

\begin{equation}
\ket{K_{S(L)}} = \frac{\ket{K_{1(2)}} +\epsilon_{S(L)} \ket{K_{2(1)}}}{\sqrt{1+ |\epsilon_{S(L)}|^2)}}
\label{eq:cp_eigen}
\end{equation}
The Eq.~\ref{eq:cp_eigen} can be finally reformulated in basis of flavor eigenstates~\cite{DiDomenico:2007zza}:
\begin{equation}
\ket{K_{S(L)}} = \frac{1}{\sqrt{2(1+ |\epsilon_{S(L)}|^2))}} [(1+ \epsilon_{S(L)}) \ket{K_{0}} \pm (1 - \epsilon_{S(L)})\ket{\bar{K}_{0}}]
\end{equation}
The parameters $\epsilon_S$ and $\epsilon_L$ can be recombined to more explicitly denote the relations with the discrete symmetries  in oscillations:  $\varepsilon \equiv \frac{\epsilon_S + \epsilon_L}{2}$ and $\delta \equiv \frac{\epsilon_S - \epsilon_L}{2}$. It can be shown that~\cite{DiDomenico:2007zza}: $\delta \neq 0$ implies CPT violation, $\Re{\varepsilon}\neq 0$ implies T violation, while $\delta \neq 0$ or $\Re{\varepsilon}\neq 0$ implies CP violation\footnote{This formulation is equivalent to the PDG notation with p,q and z complex parameters. See e.g.  Ref.~\cite{PDG}}.

As aforementioned, in DA$\Phi$NE the neutral kaon pairs produced from the $\Phi$ meson decay, as it is the case in $e^{+}e^{-}$ collisions at the DA$\Phi$NE collider, form a quantum-mechanically entangled system. In particular, the double differential decay rate of the initial entangled kaon state into the final states $f_1$ and $f_2$ at the proper times $t_1$ and $t_2$ respectively, is the observable well suited for both QM foundations and discrete symmetries studies. Integrating out 
over $t_1 + t_2$ and introducing $\Delta t=t_1 -t_2$
one can express it in the following equation ($\Delta t > 0$):
\begin{equation}
\begin{split}
    I(f_{1}, f_{2}, \Delta t) =  & C_{12}[ |\eta_{1}|^2 e^{-\Gamma_{L}\Delta t}  + |\eta_{2}|^2 e^{-\Gamma_{S}\Delta t} \\
    & - 2 |\eta_{1} \eta_{2}| e^{\frac{\Gamma_S + \Gamma_L}{2} \Delta t} cos(\Delta m\Delta t +\phi_2 - \phi_1)], 
 \label{eq:intensityGen_1}
\end{split}
\end{equation}
where $\eta_{i} = |\eta_{i}| e^{i\phi_i} = \frac{<f_i|T|\Ks>}{<f_i|T|\Kl>}$, and $C_{12} $ is the normalization factor \mbox{$\frac{|N|^2}{2 (\Gamma_{S} +\Gamma_{L})} |<f_{1}|T|K_{S}><f_2|T|K_{S}>|^2$}.
The expression contains the exponential decay terms of $\Kl$ and $\Ks$ and the interference term governed by the $\Ks$ and $\Kl$ mass difference $\Delta m = m_L - m_S$. The $\Delta m$ corresponds to the flavour oscillation frequency and is about half of the $\Ks$ decay width. This relation describes interferometric phenomena that can be exploited experimentally.
The comparison of the theoretical predictions with the experimental $\Delta t$ distribution for chosen final states can be used to determine the oscillation and decay parameters, and to perform sensitive tests of CP and CPT symmetries, as well as of the quantum decoherence.
It is worth to note that the quantum decoherence can also induce the CPT symmetry violation~\cite{Bernabeu-cpt}. More specifically, the CPT operator can become ill-defined. This scenario can be experimentally tested with neutral kaons assuming e.g the "$\omega$-effect" parametrization. For details see e.g. Ref.~\cite{Bernabeu-cpt2}. This approach can be contrasted with the CPT symmetry violation in oscillation which not necessarily leads to the quantum decoherence.

\subsection{CPT and Lorentz symmetry tests}
The CPT theorem~\cite{luders} states that any  quantum field theory that fulfills several commonly accepted requirements such as locality, Lorentz invariance, spin-statistics relation etc., must be invariant to the combined CPT transformation. All particle interaction theories that form the SM belong to that category. 

Possible violation of CPT symmetry can be attributed to effects at the Planck scale, where physical interactions are believed to be dominated by quantum gravity (QG). 
In particular, QG entails a fundamental CPT non-invariance in the environment of microscopic black-hole background \cite{wald,hawking} leading to the manifestation of the Planck-scale CPT asymmetry as a slight dissipative effect in the energy scale attainable in current experiments. As suggested in Ref. \cite{kostelecky_1}, QG effects should modify the Lagrangian of the SM of interactions by adding a small CPT-violating term and thus inducing also the Lorentz invariance breakdown. This idea led to the formulation of the Standard Model Extension (SME)~\cite{kostelecky_2}, a general framework for testing the CPT violation effects in a large spectrum of experiments.

KLOE obtained the most precise results on the CPT-violating parameter $\delta$,  as well as SME coupling constants of the SM field to a hypothetical Lorentz-violating and CPT  non-invariant background field~\cite{Kostelecky:2001ff}, based on the analysis of the $\phi\to\Ks\Kl\to(\pi^+\pi^-)(\pi^+\pi^-)$ process.
In the SME framework  the CPT violation manifests itself indirectly via Lorentz symmetry violation effects, namely the observables become dependent on the meson momentum with respect to the distant stars. 
The KLOE results reached a sensitivity at the level of $10^{-18}$ GeV~\cite{Babusci:2013gda}, which is several orders of magnitude more precise than the results obtained with other neutral meson systems~\cite{LHCb:CPT,FOCUS:CPT, Babar:CPT, D0:CPT}~\footnote{This 
precision comes close to the Planck scale of $10^{-19}$ GeV.}. The results were
obtained with a data sample corresponding to 1.7 fb$^{-1}$ integrated luminosity, and are mainly limited by the statistical errors.

\subsection{Tests of quantum decoherence}

The system of entangled kaon pair is also well suited for studies of the Einstein-Rosen-Podolsky-like (EPR) correlation and in particular to the tests of postulated
spontaneous factorization of the kaon wave-function  known as Furry hypothesis. This quantum decoherence effect can be parametrized phenomenologically by the introduction of
the effective decoherence parameter $\zeta$ into the double decay rate for a pair of neutral kaons decaying into the $\phi\to\Ks\Kl\to(\pi^+\pi^-)(\pi^+\pi^-)$ final state.
\begin{equation}
I(\pi^{+}\pi^{-}, \pi^{+}\pi^{-}, \Delta t) \propto e^{-\Gamma_{L}\Delta t}  + e^{-\Gamma_{S}\Delta t} - 2(1-\zeta_{SL})  e^{\frac{\Gamma_S + \Gamma_L}{2} \Delta t} cos(\Delta m\Delta t), 
 \label{eq:decoherence}
\end{equation}

The value $\zeta_{SL} = 1$ corresponds to the full decoherence hypothesis, while $\zeta_{SL} = 0$ is the full correlated case.
In general, the definition of $\zeta$ depends on the basis in which the initial state is expressed, because the interference term will be differently defined in $\ket{K_{S}},\ket{K_{L}}$ and in $\ket{K_{0}},\ket{\bar{K}_{0}}$ basis~\cite{decoherence}. We denote the decoherence parameters as $\zeta_{SL}$ and $\zeta_{0\bar{0}}$ respectively.

\begin{figure}[htb!]
  \centerline{\includegraphics[width=8cm]{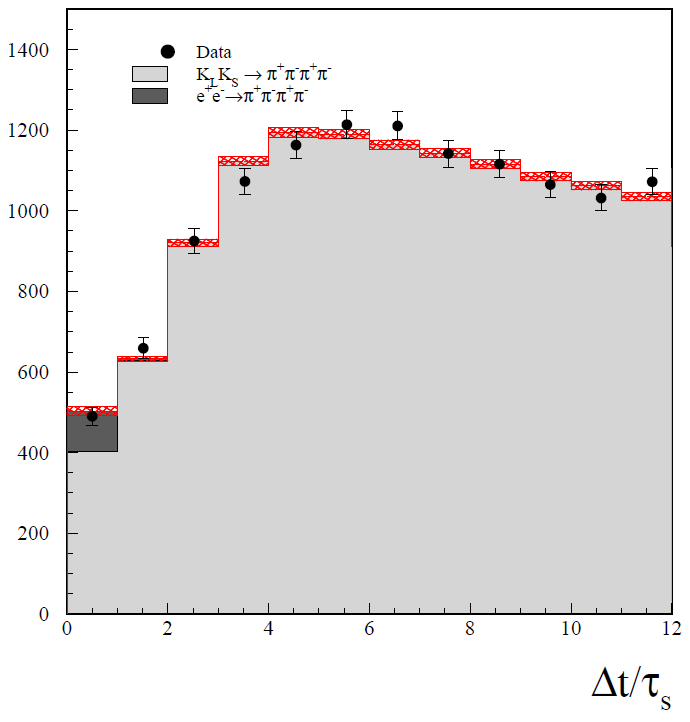}}
  \caption{The measured $\Delta t$ distribution of the time difference of the two kaon decays to $\pi^{+}\pi^{-}$. The black points denote data while the solid histogram corresponds to the fit to $\Delta t$ distribution. The bin size corresponds to time resolution of $\sigma(t) \approx \tau_{K_{S}}$. The contribution from non-resonant $e^{+} e^{-}\rightarrow  \pi^{+}\pi^{-}, \pi^{+}\pi^{-}$ events  is shown as dark grey.  Figure adapted from Ref.~\cite{decoherence}.}
  \label{fig:decoherence}
\end{figure}

The KLOE analysis based on a data sample corresponding to an integrated luminosity of $L \approx 380 \,\rm pb^{-1}$~\cite{decoherence0}, and further to the extended sample of $L \approx 1,5 \,\rm fb^{-1}$~\cite{decoherence}   yielded no observation of the decoherence. 
The measured $\Delta t$ distribution of the time difference of the decays of the $K_{S}$ and $K_{L}$ to $\pi^{+}\pi^{-}$ final states is presented in the Figure~\ref{fig:decoherence}.
The spectrum was obtained after the correction for resolution and detection efficiency effects, and taking into account the background due to the $\Ks$-regeneration on the beam pipe wall, and other backgrounds and sources of decoherence~\cite{decoherence}.

The determined $\zeta$ parameters:
\begin{equation}
  \zeta_{SL} = (0.3 \pm 1.8_{\textrm{stat}} \pm 0.6_{\textrm{syst}}) \times 10^{-2}
\end{equation}
\begin{equation}
\zeta_{0\bar{0}} = (1.4 \pm 9.5_{\textrm{stat}} \pm 3.8_{\textrm{syst}}) \times 10^{-7}
\end{equation}
are compatible with the standard QM with no decoherence effects visible. The $\zeta_{0\bar{0}}$ result improved by five orders of magnitude the previous limit~\cite{bertelmann}.

\subsection{Test of the CPT symmetry in semi-leptonic decays}
 
The relations between the charge asymmetries of decay rates 
into the two $CP$ conjugated semi-leptonic final states,
$\pi^{-} e^{+} \nu$ and $\pi^{+} e^{-} \bar{\nu}$ can be expressed in terms of the CP- and CPT-violating parameters.

In particular, the charge asymmetries for the physical states $\Ks$ and $\Kl$ are defined as:
\begin{equation}
    \begin{aligned}
        A_{S,L}
        & = 
        \frac{\Gamma(K_{S,L} \rightarrow \pi^{-} e^{+} \nu) - \Gamma(K_{S,L}
            \rightarrow
            \pi^{+} e^{-}
        \bar{\nu})}{\Gamma(K_{S,L} \rightarrow \pi^{-} e^{+} \nu) + \Gamma(K_{S,L}
            \rightarrow
        \pi^{+} e^{-} \bar{\nu})}   
    \end{aligned}
\end{equation}
can be approximated at the  first order in small parameters \cite{handbook_cp}:
\begin{equation}
    \begin{aligned}
        A_{S,L}   & = 2 \left[  \Re\left( \varepsilon\right) \pm \Re \left(\delta \right) - \Re (y) \pm \Re( x_{-}) \right]       
    \end{aligned}
\end{equation}
with $\Re\left(\varepsilon \right) $ and $\Re \left(\delta \right)$ implying $T$- and $CPT$-violation in the $K^0-\bar{K^0}$ mixing, respectively,
$\Re( y)$ and $\Re( x_-)$ implying $CPT$ violation in $\Delta S=\Delta Q$ and $\Delta S \neq \Delta Q$ decay amplitudes, 
respectively, and all parameters implying $CP$ violation.
If $CPT$ symmetry holds, then 
the two asymmetries are expected to be identical
$A_{S}\!=\!A_{L}\!=\!2\,\Re,(\varepsilon)\!\simeq\!3\!\times\!10^{-3}$,
each accounting for the $CP$ impurity in the mixing in the corresponding physical state.
The sum $A_{S}+A_{L}=4\left(\Re,\varepsilon-\Re,y\right)$ can be used to extract the $CPT$-violating parameter 
$\Re (y)$ once the measured value of $\Re (\varepsilon)$ is provided as input.
\par
The combinations $A_{S}\pm A_L$ (dominated by the uncertainty on $A_S$) can be used to  
 improve the semi-leptonic decay
contribution to the $CPT$ test obtained imposing the unitarity relationship,
originally derived by Bell and Steinberger \cite{BS},
and yielding the most stringent limits on $\Im(\delta)$ and the mass difference $m(K^0)-m(\bar{K^0})$~\cite{Ambrosino:2006ek,Patrignani:2016xqp}. 

 Using 1.63~fb$^{-1}$ of integrated luminosity collected by the KLOE experiment~\cite{Kisielewska-thesis}, about $7\times 10^4$ $\Ks\rightarrow\pi^{\pm}e^{\mp}\nu$ decays have been reconstructed. The measured value of the charge asymmetry for this decay is $\asresult$~\cite{daria2}, which is almost twice more precise than the previous KLOE result~\cite{Ambrosino:2006si}. The combination of these two measurements gives $\asresultcombined$ and, together with the asymmetry of the $\Kl$ semi-leptonic decay, from the KTeV Collaboration: $\alktev$~\cite{AlaviHarati:2002bb},  provides significant tests of the $CPT$ symmetry.
    
\begin{figure}[htb!]
    \centering
      \includegraphics[width=0.9\textwidth]{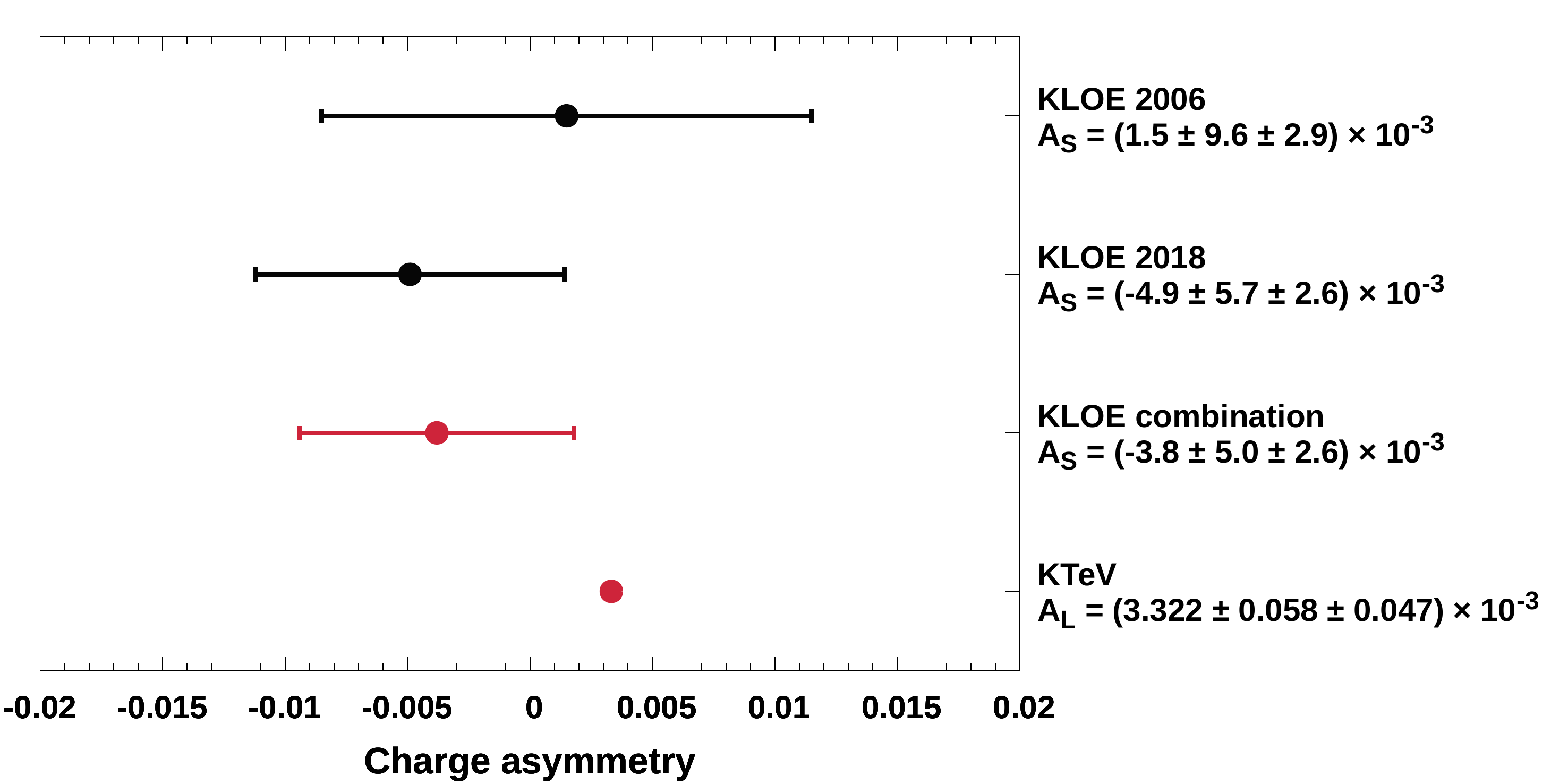}
      \caption{
Comparison of the previous 
result for $A_S$ 
(KLOE 2006~\cite{Ambrosino:2006si}), the result presented in the newest paper (KLOE 2018)~\cite{daria2} and the combination of the two.
The KTeV result for $A_L$~\cite{AlaviHarati:2002bb} is also shown.
The uncertainties of the points correspond to the  statistical and systematic uncertainties summed in quadrature.
      }
      \label{fig::plotresultsas}
\end{figure}

Using $\Re(\delta) = (2.5 \pm 2.3) \times 10^{-4}$ (Ref.~\cite{Patrignani:2016xqp})
and $\Re(\varepsilon) = (1.596 \pm 0.013) \times 10^{-3}$~(Ref.\cite{Ambrosino:2006ek})
the $CPT$ violating parameters $\Re(x_-)$ and $\Re(y)$ are extracted: 
\begin{align}
    \Re(x_-) & = (-2.0 \pm 1.4) \times 10^{-3}, \\
    \Re(y) & = (1.7 \pm 1.4) \times 10^{-3},
\end{align}
which are consistent with $CPT$ invariance. 

\subsection{T and CPT direct test in transitions of neutral kaons}
The CPT symmetry, as well as T symmetry, can be directly tested at KLOE by a new method in kaon transitions  
based on a comparison between rates of processes from given kaon flavor-defined to given CP-defined states and their time-reversal conjugates obtained by an exchange of initial and final states~\cite{Bernabeu-t,Bernabeu-cpt}. This allows for a model-independent test of T and CPT symmetries. Such a test has been performed only in the case of neutral B mesons, where the first evidence of T violation was observed~\cite{lees}. In the case of KLOE-2, it can be pursued with the $K^0-\bar{K}^0$ system~\cite{Gajos-thesis}. For this purpose,
quantum-entangled meson pairs are used to identify a particle initial state by
the decay of its entangled partner, while the final state can be tagged using the semi-leptonic and hadronic
decays into two and three pions.
Then, two T violating observables can be defined, by using the ratios between the rates of the two classes of processes $\Ks \Kl \to \pi^\pm e^\mp \nu, 3\pi^0$ and $\Ks \Kl \to \pi^+ \pi^-, \pi^\pm e^\mp \nu$:
\begin{equation}
      R_2(\Delta t) = \frac{P[K^0(0) \to K_-(\Delta t)]}{P[K_-(0) \to K^0(\Delta t)]} \sim \frac{I(\pi^+e^- \bar{\nu}, 3\pi^0;\Delta t)}
  {I(\pi^+\pi^-,\pi^-e^+\nu; \Delta t)}
 \label{eq:ratioTsymm_1}
\end{equation}
\begin{equation}
  R_4(\Delta t) = \frac{P[\bar{K}^0(0) \to K_-(\Delta t)]}{P[K_-(0) \to \bar{K}^0(\Delta t)]} \sim \frac{I(\pi^-e^+ \nu, 3\pi^0;\Delta t)}{I(\pi^+\pi^-,\pi^+e^-\bar{\nu}; \Delta t)},
\label{eq:ratioTsymm_2}
\end{equation}
where, as previously introduced, $I( f_1 , f_2 ;\Delta t)$ denotes the number of recorded events characterized by a time-ordered pair
of kaon decays $f_1$ and $f_2$ separated by an interval of proper kaon decay times $\Delta t$. Any
deviations of the asymptotic value of these ratios from unity for large transition times would
be a manifestation of T violation.

This test can be extended to test CPT symmetry through the determination of the asymptotic level of the double ratio defined as:
\begin{equation}
\begin{split}
  \frac{R_2^{CPT}}{R_4^{CPT}} &   
   = \frac{\mathrm{P}[\kaon(0)\to\Km(\Delta t)] / \mathrm{P}[\Km(0)\to\akaon(\Delta t)]}{\mathrm{P}[\akaon\to\Km(\Delta t)] / \mathrm{P}[\Km(0)\to\kaon(\Delta t)]} \\
   & \stackrel{\Delta t \gg \tau_S}{=} 1 - 8\Re(\delta) -8\Re(x_{-})  
  \label{double_ratio_CPT},
\end{split}
\end{equation}
where the $\Re(\delta)$ and $\Re(x_{-})$ are parameters violating the CPT symmetry in $\kaon\akaon$ mixing and the $\Delta S=\Delta Q$ rule, respectively, already introduced in the previous section. This double ratio represents a robust CPT-violation sensitive observable~\cite{Bernabeu-t,Bernabeu-cpt} which has never been measured to date.
A percent level accuracy has been obtained on the double ratio measurement with 1.7 fb$^{-1}$ KLOE data sample~\cite{DiCicco, Gajos:2018psd}.

\section{Outlook and prospects}
\label{sec:outlook}
The article presents a summary of the most recent results of the KLOE-2 Collaboration in hadron physics, dark matter searches, discrete symmetries and decoherence tests with kaons. Most of the presented results are limited by the statistical errors.

The ongoing analyses are taking advantage not only of the larger statistics but of the information provided by the new detectors, which are expected to improve significantly the key experimental parameters e.g. time resolution, or acceptance coverage and consequently lead to smaller systematical uncertainties.

In the near future, the new results based on the full collected statistics are expected.
For example, the uncertainty on A$_S$, should be reduced to the level of $\sim$ 3$\times$10$^{-3}$ and the accuracy in the double ratio measurement for the CPT test to the 10$^{-3}$ level of precision. The results on the U boson should lower the upper limit by roughly a factor of 2 or better. Also, the opening for new searches in different models of dark matter mediators is being investigated.

\section*{Acknowledgments}

We warmly thank our former KLOE colleagues for the access to the data collected during the KLOE data taking campaign.
We thank the DA$\Phi$NE team for their efforts in maintaining low background running conditions and their collaboration during all data taking. We want to thank our technical staff: 
G.F. Fortugno and F. Sborzacchi for their dedication in ensuring efficient operation of the KLOE computing facilities; 
M. Anelli for his continuous attention to the gas system and detector safety; 
A. Balla, M. Gatta, G. Corradi and G. Papalino for electronics maintenance; 
C. Piscitelli for his help during major maintenance periods. 
This work was supported in part 
by the Polish National Science Centre through the Grants No.\
2013/11/B/ST2/04245,
2014/14/E/ST2/00262,
2014/12/S/ST2/00459,
2016/21/N/ST2/01727,
2016/23/N/ST2/01293,
2017/26/M/ST2/00697.




\printbibliography

@article{physics_kloe2,
    author = "G. Amelino-Camelia et al.",
    journal = "Eur. Phys. J.",
    volume = "\textbf{C 68}",
    pages = "619-681",
    year = 2010
    }

@article{DAFNE-KLOE,
    author = "A.~Gallo et al.",
    journal = "Conf.Proc.",
    volume ="\textbf{C060626}",
    pages = "604-606",
    year = "2006"
}

@article{DAFNE-KLOE2,
    author = "C.~Milardi et al.",
    journal = "JINST",
    volume ="\textbf{7}",
    pages = "T03002",
    year = "2012"
}

@InProceedings{Milardi:eeFACT2018-MOYAA02,
  author       = {C.~Milardi et al.},
  
  booktitle    = {Proc. 62nd ICFA ABDW on High Luminosity Circular e$^+$e$^-$ Colliders (eeFACT'18),
                  Hong Kong, China, 24-27 September 2018},
  pages        = {23--29},
  series       = {ICFA ABDW on High Luminosity Circular e$^+$e$^-$ Colliders},
  number       = {62},
  publisher    = {JACoW Publishing},
  year         = {2019}
}

@article{KLOE2_proposal,
    author = "G.~Amelino-Camelia et al.",
    journal = "Eur. Phys. J",
    volume ="{\textbf{C68}}",
    pages = "619-681",
    year = "2010"
}

@online{CW1,
    author = "P. Raimondi and D. Shatilov and M. Zobov",
    archivePrefix = "arXiv",
    eprint = "physics/0702033",
    note = "LNF-07-003-IR",
    year = "2007"
}

@article{CW2,
    author = "P. Raimondi and D. Shatilov and M. Zobov",
    journal = "Proc. 11th European Particle Accelerator Conference (EPAC 2008)",
    volume ="",
    pages = "2620-2622",
    year = "2008"
}

@article{DAFNE-2,
    author = "M. Zobov et al.",
    journal = "Phys. Rev. Lett.",
    volume = "\textbf{104}",
    pages = "174801",
    year = "2010"
}

@article{siddharta,
    author = "SIDDHARTA Collaboration",
    journal = "Eur. J. Phys.",
    volume = "\textbf{A31}",
    pages = "537",
    year = "2007"
}

@article{siddharta2,
    author = "SIDDHARTA-2 Collaboration",
    journal = "EPJ Web of Conferences",
    volume = "\textbf{181}",
    pages = "01004",
    year = "2018"
}

@article{nuovocimento,
    author = "F. Bossi and E. De Lucia and J. Lee-Franzini and S. Miscetti and M. Palutan and KLOE Collaboration",
    journal = "Rivista del Nuovo Cimento",
    volume = "\textbf{vol. 031 Issue 10}",
    pages = "531-623",
    year = "2008"
}

@article{driftchamber,
    author = "M. Adinolfi et al.",
    journal = "Nucl. Instrum. Meth.",
    volume = "\textbf{A488}",
    pages = "51-73",
    year = "2002"
}

@article{calorimeter,
    author = "M. Adinolfi et al.",
    journal = "Nucl. Instrum. Meth.",
    volume = "\textbf{A482}",
    pages = "364-386",
    year = "2002"
}

@article{cgem-it,
    author = "A. Balla et al.",
    journal = "Nucl. Instrum. Meth.",
    volume = "\textbf{A732}",
    pages = "221",
    year = "2013"
}

@article{HET-LET,
    author = "D. Babusci et al.",
    journal = "Acta Phys. Pol.",
    volume = "\textbf{B46}",
    pages = "81",
    year = "2015"
}

@article{QCALT,
    author = "A. Balla et al.",
    journal = "Nucl. Instrum. Meth.",
    volume = "\textbf{A718}",
    pages = "95-96",
    year = "2013"
}

@article{CCALT,
    author = "M. Cordelli et al.",
    journal = "Nucl. Instrum. Meth.",
    volume = "\textbf{A718}",
    pages = "81-81",
    year = "2013"
}

@article{arbuzov,
    author = "A. B. Arbuzov and D. Haidt and C. Matteuzzi and M. Paganoni and L. Trentadue",
    journal = "Eur. Phys. J.",
    volume = "\textbf{C34}",
    pages = "267",
    year = "2004"
}

@article{abbiendi,
    author = "G. Abiendi et al. [OPAL Collaboration]",
    journal = "Eur. Phys J.",
    volume = "\textbf{C45}",
    pages = "1",
    year = "2006"
    }

@article{acciarri,
    author = "M. Acciarri et al. [L3 Collaboration]",
    journal = "Phys. Lett.",
    volume = "\textbf{B476}",
    pages = "40",
    year = "2000"
    }

@article{odaka,
    author = "S. Odaka et al. [VENUS Collaboration]",
    journal = "Phys. Rev. Lett.",
    volume = "\textbf{81}",
    pages = "2428",
    year = "1998"
    }

@article{levine,
    author = "I. Levine et al. [TOPAZ Collaboration]",
    journal = "Phys. Rev. Lett.",
    volume = "\textbf{78}",
    pages = "424",
    year = "1997"
}

@article{kallen,
    author = "G. K{\"a}ll{\`e}n and A. Sabry",
    journal = "K. Dan. Vidensk. Selsk. Mat.-Fys. Medd.",
    volume = "29",
    pages = "17",
    year = "1955"
}

@article{steinhausen,
    author = "M. Steinhauser",
    journal = "Phys. Lett.",
    volume = "B429",
    pages = "158",
    year = "1998"
}

@article{aem-1,
    author = "A. Anastasi et al.",
    journal = "Phys. Lett.",
    volume = "\textbf{B767}",
    pages = "485-492",
    year = "2017"
}

@article{aem_theory,
    author = "F. Jegerlehner",
    journal = "Nuovo Cim.",
    volume = "\textbf{C034S1}",
    pages = "31-40",
    year = "2011"
}

@article{PDG,
    author = "K. A. Olive et al.",
    journal = "Chin. Phys.",
    volume = "\textbf{C38}",
    pages = "090001",
    year = "2014"
}

@article{2picross,
    author = "D. Babusci et al.",
    journal = "Phys. Lett.",
    volume = "\textbf{B720}",
    pages = "336-343",
    year = "2013"
}

@article{G-S,
    author = "G. J. Gounaris and J. J. Sakurai",
    journal = "Phys. Rev. Lett.",
    volume = "\textbf{21}",
    pages = "244",
    year = "1968"
}

@article{ChPT,
    author = "Ambrosino et al.",
    journal = "JHEP",
    volume = "\textbf{0805}",
    pages = "006",
    year = "2008"
}

@article{KLOEdalitz_16,
    author = "A. Anastasi et al.",
    journal = "JHEP",
    volume = "\textbf{05}",
    pages = "019",
    year = "2016"
}

@article{KLOEdalitz_08,
    author = "F. Ambrosino et al.",
    journal = "JHEP",
    volume = "\textbf{05}",
    pages = "006",
    year = "2008"
}

@article{wasa_dalitz,
    author = "P. Adlarson et al.",
    journal = "Phys. Rev.",
    volume = "\textbf{C90}",
    pages = "045207",
    year = "2014"
}

@article{bes_dalitz,
    author = " M. Ablikim et al.",
    journal = "Phys. Rev.",
    volume = "\textbf{D92}",
    pages = "012014",
    year = "2015"
}

@article{integral,
    author = "P. Jean et al.",
    journal = "Astrophys.",
    volume = "{\bf{407}}",
    pages = "L55",
    year = "2003"
}

@article{cogent,
    author = "C. E. Aalseth et al.",
    journal = "Phys. Rev. Lett.",
    volume = "{\bf{107}}",
    pages = "141301",
    year = "2011"
}

@article{dama,
    author = "R. Bernabei et al.",
    journal = "Int. J. Mod. Phys.",
    volume = "{\bf{D13}}",
    pages = "2127",
    year = "2004"
}

@article{libra,
    author = "R. Bernabei et al.",
    journal = "Eur. Phys. J.",
    volume = "\bf{C56}",
    pages = "333",
    year = "2008"
}

@article{atic,
    author = "J. Chang et al.",
    journal = "Nature",
    volume = "\textbf{456}",
    pages = "362",
    year = "2008"
}

@article{hess1,
    author = "F. Aharonian et al.",
    journal = "Phys. Rev. Lett.",
    volume = "{\bf{101}}",
    pages = "261104",
    year = "2008"
}

@article{hess2,
    author = "F. Aharonian et al.",
    journal = "Astron. Astrophys.",
    volume = "{\bf{508}}",
    pages = "561",
    year = "2009"
}

@article{fermi,
    author = "A. A. Abdo et al.",
    journal = "Phys. Rev. Lett.",
    volume = "{\bf{102}}",
    pages = "181101",
    year = "2009"
}

@article{ext1,
    author = "B. Holdom",
    journal = "Phys. Lett.",
    volume = "{\bf{B166}}",
    pages = "196",
    year = "1985"
}

@article{ext2,
    author = "C. Boehm and, P. Fayet",
    journal = "Nucl. Phys.",
    volume = "{\bf{B683}}",
    pages = "219",
    year = "2004"
}

@article{ext3,
    author = "P. Fayet",
    journal = "Phys. Rev.",
    volume = "{\bf{D75}}",
    pages = "115017",
    year = "2007"
}

@article{ext4,
    author = "M. Pospelov and A. Ritz and M.B. Voloshin",
    journal = "Phys. Lett.",
    volume = "{\bf{B662}}",
    pages = "53",
    year = "2008"
}

@article{ext5,
    author =  "Y. Mambrini",
    journal = "J. Cosmol. Astropart. Phys.",
    volume = "{\bf{1009}}",
    pages = "022",
    year = "2010"
}

@article{KLOE_UL1,
    author = "F. Archilli et al.",
    journal = "Phys. Lett.",
    volume = "\textbf{B706}",
    pages = "251",
    year = "2012"
}

@article{KLOE_UL2,
    author = "D. Babusci et al.",
    journal = "Phys. Lett.",
    volume = "\textbf{B720}",
    pages = "111",
    year = "2013"
}

@article{enrico,
    author = "D. Babusci et al.",
    journal = "Phys. Lett.",
    volume = "\textbf{B747}",
    pages = "365",
    year = "2015"
}

@article{Batell,
    author = "B. Batell and M. Pospelov and A. Ritz",
    journal = "Phys. Rev.",
    volume = "\textbf{D79}",
    pages = "115008",
    year = "2009"
}

@article{eeg,
    author = "A. Anastasi et al.",
    journal = "Phys. Lett.",
    volume = "\textbf{B750}",
    pages = "633",
    year = "2015"
}

@article{mmg,
    author = "D. Babusci et al.",
    journal = "Phys. Lett.",
    volume = "\textbf{B736}",
    pages = "459",
    year = "2014"
}

@article{ppg,
    author = "A. Anastasi et al.",
    journal = "Phys. Lett",
    volume = "\textbf{B757}",
    pages = "356",
    year = "2016"
}

@article{babar,
    author = "J.P. Lees et al.",
    journal = "Phys. Rev. Lett.",
    volume = "\textbf{108}",
    pages = "211801",
    year = "2012"
}

@article{na48,
    author = "J.R. Batley et al.",
    journal = "Phys. Lett.",
    volume = "\textbf{B746}",
    pages = "178185",
    year = "2015"
}

@article{lhcb,
    author = "S. Abrahamyan et al.",
    journal = "Phys. Rev. Lett.",
    volume = "\textbf{107}",
    pages = "191804",
    year = "2011"
}

@article{Bernabeu-t,
    author = "J.~Bernabeu et al.",
    journal = "Nucl. Phys.",
    volume = "\textbf{B868}",
    pages = "102-119",
    year = "2013"
}

@article{Bernabeu-cpt,
    author = "J.~Bernabeu et al.",
    journal = "JHEP",
    volume = "\textbf{10}",
    pages = "139",
    year = "2015"
}

@article{lees,
    author = "J. P. Lees et al.",
    journal = "Phys. Rev. Lett.",
    volume = "\textbf{109}",
    pages = "211801",
    year = "2012"
}

@article{Babusci:2013gda,
    author = "D.~Babusci et al.",
    journal = "Phys. Lett.",
    volume = "\textbf{B730}",
    pages = "89-94",
    year = "2014"
}

@article{Kostelecky:2001ff,
    author =  "V.A.~Kostelecky",
    journal = "Phys. Rev.",
    volume = "\textbf{D64}",
    pages = "076001",
    year = "2001"
}

@article{decoherence0,
    author = "F.~Ambrosino et al.",
    journal = "Phys. Lett.",
    volume = "\textbf{B642}",
    pages = "315",
    year = "2006"
}

@article{decoherence,
    author = "A.~Di~Domenico",
    journal = "Journal of Physics: Conf. Ser.",
    volume = "\textbf{171}(1)",
    pages = "012008",
    year = "2009"
}

@article{bertelmann,
    author = "R.~.A.~Bertlmann and W.~Grimus and B.~C.~Hiesmayr",
    journal = "Phys. Rev.",
    volume = "\textbf{D60}",
    pages = "114032",
    year = "1999"
}

@article{daria2,
    author = "A.~Anastasi et al.",
    journal = "JHEP",
    volume = "\textbf{21}",
    pages = "9",
    year = "2018"
}

@article{wald,
    author = "R.M. Wald",
    journal = "Phys. Rev.",
    volume = "\textbf{D21}",
    pages = "2742",
    year = "1980"
}

@article{hawking,
    author = "S.W. Hawking",
    journal = "Comm. Math. Phys.",
    volume = "\textbf{87}",
    pages = "395",
    year = "1982"
}

@article{kostelecky_1,
    author = "V.A. Kostelecky and R. Potting",
    journal = "Phys. Lett.",
    volume = "\textbf{B 381}",
    pages = "89",
    year = "1996"
}

@article{kostelecky_2,
    author = "B.~R. Edwards and V.~A. Kostelecky",
    journal = "Phys. Lett.",
    volume = "\textbf{B795}",
    pages = "620-626",
    year = "2019"
}

@book{handbook_cp,
  author    = "L. Maiani", 
  title     = "CP and CPT violation in neutral kaon decays",
  publisher = "The Second DA$\phi$NE Physics Handbook",
  volume    = "I",
  pages    = "3",
  address   = "INFN-LNF, Frascati",
  year     = "1995"
}

@book{cp_sozzi,
  author    = "M.~S.~Sozzi", 
  title     = "Discrete symmetries and CP violation",
  publisher = "Oxford University Press",
  volume    = "1",
  year     = "2008"
}

@article{luders,
    author = "G.~L{\"u}ders",
    journal = "Ann. Phys.",
    volume = "2",
    pages = "1",
    year = "1957"
}

@article{BS,
    author = "J.~S. Bell and J.~Steinberger",
    journal = "Proc. of the Oxford Int. Conf. on Elementary Particles",
    volume = "",
    pages = "",
    year = "1965"
}

@article{Ambrosino:2006ek,
    author = "{F.~Ambrosino et~al. [\scshape {KLOE}} Collaboration]",
    journal = "JHEP",
    volume = "{\bfseries 12}",
    pages = "11",
    year = "2006"
}

@article{Patrignani:2016xqp,
    author = "C.~Patrignani et al. (Particle Data Group)",
    journal = "Chin. Phys. C",
    volume = "{\bfseries 40}",
    pages = "100001",
    year = "2016"
}

@article{AlaviHarati:2002bb,
    author = "A.~Alavi-Harati et~al. [{KTeV} Collaboration]",
    journal = "Phys. Rev. Lett.",
    volume = "{\bfseries 88}",
    pages = "181601",
    year = "2002"
}

@book{DiDomenico:2007zza,
    author = "A. Di Domenico",
    title = "Handbook on Neutral Kaon Interferometry at a $\phi$-factory",
    series = "Frascati Physics Series",
    volume = "43",
    year = "2007"
}

@article{Ambrosino:2006si,
    author = "F.~Ambrosino et~al. [\scshape {KLOE} Collaboration]",
    journal = "Phys. Lett.",
    volume = "{\bfseries B 636}",
    pages = "173",
    tear = "2006"
}

@online{PDG_CP,
    url = "http://pdg.lbl.gov/2019/reviews/rpp2018-rev-cp-violation.pdf"
}

@article{DiCicco,
    author = "A.~Di Cicco",
    journal = "EPJ Web Conf.",
    volume = "212",
    pages = "09001",
    year = "2019"
}

@article{Gajos:2018psd,
      author     = "A.~Gajos",
      journal        = "Hyperfine Interact.",
      volume         = "239",
      pages          = "53",
      year           = "2018"
}

@article{LHCb:CPT,
      author     = "R.~Aaij et al. [LHCb Collaboration]",
      journal        = "Phys. Rev. Lett.",
      volume         = "{\bfseries 116}",
      pages          = "241601",
      year           = "2016"
}

@article{D0:CPT,
      author     = "V.M. Abazov et al. [D0 Collaboration]",
      journal        = "Phys. Rev. Lett.",
      volume         = "{\bfseries 115}",
      pages          = "161601",
      year           = "2015"
}

@article{FOCUS:CPT,
      author     = "J. Link et al. [FOCUS Collaboration]",
      journal        = "Phys. Lett. B",
      volume         = "{\bfseries 556}",
      pages          = "7",
      year           = "2003"
}

@article{Babar:CPT,
      author     = "B. Aubert et al. [BaBar Collaboration]",
      journal        = "Phys. Rev. Lett.",
      volume         = "{\bfseries 100}",
      pages          = "131802",
      year           = "2008"
}

@article{Bernabeu-cpt2,
      author     = "J.~Bernabeu, N.~E.~Mavromatos, S.~Sarkan",
      journal        = "Phys.Rev. D",
      volume         = "{\bfseries 74}",
      pages          = "045014",
      year           = "2006"
}

@thesis{Gajos-thesis,
      author     = "A.~Gajos",
      title        = "Investigations of fundamental symmetries with the electron-positron systems",
      school          = "Jagiellonian University",
      year           = "2018"
}

@thesis{Kisielewska-thesis,
      author     = "D.~Kisielewska",
      title      = "Studies of CPT symmetry violation in matter-antimatter systems",
      school          = "Jagiellonian University",
      year           = "2019"
}
\end{document}